\documentclass[preprint2]{aastex}

\usepackage[hidelinks]{hyperref}

\shorttitle{Solar wave-field simulation for testing deep meridional flow measurements}
\shortauthors{Hartlep et al.}

\usepackage{color}                       

\newcommand{\degree}{\ensuremath{^\circ}}

\begin{document}


\title{Solar wave-field simulation for testing prospects of helioseismic measurements of deep meridional flows}


%

\author{T. Hartlep, J. Zhao, and A.G. Kosovichev}
\affil{W.W.~Hansen Experimental Physics Laboratory, Stanford University, Stanford, CA, USA}

\and
 
\author{ N.N.~Mansour}
\affil{NASA Ames Research Center, Moffett Field, CA, USA}




\begin{abstract}

The meridional flow in the Sun is an axisymmetric flow that is generally poleward directed at the surface, and is presumed to be of fundamental importance in the generation and transport of magnetic fields.
Its true shape and strength, however, is debated.
We present a numerical simulation of helioseismic wave propagation in the whole solar interior in the presence of a prescribed, stationary, single-cell, deep meridional circulation serving as a test-bed for helioseismic measurement techniques.
A deep-focusing time-distance helioseismology technique is applied to the artificial data showing that it can in fact be used to measure the effects of the meridional flow very deep in the solar convection zone.
It is shown that the ray-approximation which is commonly used for interpretation of helioseismology measurements remains a reasonable approximation even for the very long distances between $12\degree$ and $42\degree$ corresponding to depths between 52 and 195~Mm considered here.
From the measurement noise we extrapolate that on the order of a full solar cycle may be needed to probe the flow all the way to the base of the convection zone. 

\end{abstract}






\section{Motivation and Objectives}

The meridional circulation is known as a poleward flow near the solar surface and is believed to be an important part of the dynamo process in the solar convection zone.
Its true strength and shape in the deeper interior is presently unknown or at least highly debated.
There are conflicting theories and observational evidences.
On the one hand, theoretical works \citep[e.g.,][]{2005AN....326..379K} usually favor deep meridional flows.
Due to mass conservation, the return flow then is relatively strong with flow velocities of the similar same order as the flow near the surface.

On the other hand, observational evidence points towards a shallow meridional flow.
\citet{2007AN....328.1009M} measured {\it p}-mode frequency shifts from three months of SOHO/MDI data and inferred a meridional flow that reverses at a depth of about 40~Mm.
They also find evidence of a possible second reversal deeper below.
\citet{2011arXiv1103.1561H} uses the advection of convection cells by the meridional circulation to measure the flow velocity at depth and finds a return flow starting at a depth of only 35~Mm.
Other techniques, such as time-distance helioseismology have also been used to measure the meridional flow~\citep[e.g.,][]{1998ESASP.418..775G,2001ApJ...559L.175C,2004ApJ...603..776Z,2012SoPh..275..375Z} but primarily for small depths.

The goal of this work was to numerically simulate the propagation of helioseismic waves in the Sun in the presence of meridional and other large scale flows, and to use the artificial data from such simulations to evaluate the possibility of measuring small flows very deep in the Sun, in particular using time-distance helioseismology.
There have been previous works that addressed the measurability of deep meridional flows in the Sun such as~\citet{2008ApJ...689L.161B}.
They concluded that as much observations as a full solar cycle may be needed to resolved flows near the base of the solar convection zone using the helioseismic holography technique. 
Various models of meridional flows have been proposed, and for this paper we have performed a numerical simulation of the solar acoustic wave field for a global-Sun model which included a meridional circulation with a deep return flow, and have performed time-distance helioseismology measurements for this model as if it were observational data.
We compare the measurements with the predictions of the ray-path helioseismology theory and estimate the noise level due to the stochastic nature of solar oscillations, and the sensitivity of the helioseismology technique.

\section{Numerical Simulation}

\subsection{Simulation Code}

We have built a numerical code that solves the linearized propagation of helioseismic waves throughout the entire solar interior in the presence of a background structure and flow model~\citep{har05}.
This code has been used in previous studies to simulate the effects of localized sound speed perturbations, e.g., for testing helioseismic far-side imaging by simulating the effects of model sunspots on the acoustic field~\citep{2008ApJ...689.1373H,2009SoPh..258..181I}, for validating time\,--\,distance helioseismic measurements of tachocline perturbations~\citep{2009ApJ...702.1150Z}, and for studying the effects of localized subsurface perturbations \citep{2011SoPh..268..321H}.
For the present case, the code has been extended to include the effects that mass flows have on the propagation of helioseismic waves.

The simulation code models solar acoustic oscillations in a spherical domain using the Euler equations linearized around a stationary background state characterized by the background density, $\rho_0$, mass flows velocity, $\mathbf{v_0}$, sound speed, $c_0$, and acceleration due to gravity, $\mathbf{g_0}$.
The equations derived for the perturbations around the base state are then:
\begin{eqnarray}
  \label{Eq:E1}
  \partial_t \rho^\prime & = & - \nabla\cdot\mathbf{m}^\prime + \cal{S},  \\
  \label{Eq:E2}
  \partial_t \mathbf{m}^\prime & = & - \nabla c_0^2 \rho^\prime + \rho^\prime \mathbf{g_0} + \mathbf{v_0} (\mathbf{v_0}\cdot\nabla\rho^\prime)   \\
  & + & \rho^\prime ( \mathbf{v_0}\cdot\nabla\mathbf{v_0} + \mathbf{v_0} \nabla\cdot\mathbf{v_0} ) - (\mathbf{v_0}\cdot\nabla\mathbf{m}^\prime  \nonumber \\
  & + & \mathbf{m}^\prime\cdot\nabla\mathbf{v_0} + \mathbf{m}^\prime\nabla\cdot\mathbf{v_0} + \mathbf{v_0}\nabla\cdot\mathbf{m}^\prime), \nonumber 
\end{eqnarray}
where $\rho^\prime$ and $\mathbf{m}^\prime=\rho^\prime \mathbf{v}_0 + \rho_0 \mathbf{v}^\prime$ are the density and momentum perturbations associated with the
waves, respectively. 

Several simplifications where used in deriving these equations.
In particular, perturbations of the gravitational potential have been neglected, and the adiabatic approximation has been used.
The entropy gradient of the background model has been neglected in order to make the linearized equations convectively stable.
Previous calculations have shown that this assumption does not significantly change the propagation properties of acoustic waves including their frequencies, except for the acoustic cut-off frequency, which is slightly reduced.
Because of this simplification, no separate energy equation needs to be solved.


The equations are formulated in a non-rotating frame.
Rotation can however be accounted for by prescribing an appropriate flow.
This approach saves computing the usual Coriolis and centrifugal forces that appear in the equations written in a rotating frame.
There is no additional time-stepping or stability constraint on the numerical method because the rotation speed in the solar interior is always significantly smaller than the speed of sound.

In the Sun, vigorous turbulent flows near the photosphere are the primary sources of acoustic perturbations.
This however is a non-linear process and is lost by linearizing the equations.
A random function $\cal{S}$ has therefore been added to Eqn.~\ref{Eq:E1} to mimic the excitation of these acoustic perturbations.
$\cal{S}$ is random in time and horizontal space, and peaks at a depth of 150 km below the photosphere.

The equations are solved in spherical coordinates using a pseudo-spectral method.
2/3-dealiasing is used.
Scalar quantities such as pressure and density are expanded in spherical harmonic basis functions for their angular structure and B-splines \citep{DeBoor87,Loulou97,Hartlep04_3,har05} for their radial dependence.
Vector fields such as the perturbation of the momentum are expanded in vector spherical harmonics and B-splines. 
Vector spherical harmonics \citep{1935PhRv...47..139H,Hill54} are analogous to spherical harmonic functions but for vector quantities.
They were selected here because the coordinate singularities in spherical coordinates can be treated straightforwardly in these basis functions.
Due to the nature of spherical harmonics, no special treatment is necessary for the poles, and only the center of the sphere needs special treatment.
For the expanded variables to be finite at the center, we must enforce that the expansion coefficients follow specific asymptotic behaviors as $r\to0$. 
For details on how this is implemented using B-splines, the reader is referred to \cite{Hartlep_CTR_SUMMER_2006}.

B-splines of polynomial order 4 are used in our simulations.
The spacing of the generating knot points is chosen to be proportional to the local speed of sound. 
This results in a high radial resolution near the solar surface, where the sound speed is low (less than 7~km/s), and a significantly lower resolution in the deep interior, where the sound speed surpasses 500~km/s.
This provides a constant Courant-Friedrichs-Lewy (CFL) condition throughout the domain.
A total of 350 B-splines are used to discretize the entire simulation domain reaching from the solar center to an outer radius of 700 Mm.

Effectively non-reflecting boundary conditions are used at the upper boundary by means of an absorbing buffer layer by adding terms $-\sigma \rho^\prime$ and $-\sigma\mathbf{m}^\prime$ in the above equations.
The equations are recast using an integrating factor $\exp{\sigma t}$, and advanced using a staggered leapfrog scheme in which $\rho^\prime$ and $\mathbf{m}^\prime$ are offset by half of a time step.
The purpose of the buffer layer is to damp waves passing thought the temperature minimum before they reach the numerical boundary.
Such waves would ordinarily escape into the chromosphere, and we do not want them to artificially reflect back from the numerical boundary. 
The damping coefficient $\sigma$ is non-negative, smooth and constant in time.
It is zero in the interior and increases smoothly into the buffer layer.
Similar damping is used in the deepest interior near the solar center.
Waves of high spherical harmonic degree do not travel very deep.
Their lower turning radius $r_t$ given by
\begin{equation}
 c_0^2(r_t)/r_t^2=\omega^2/l(l+1),
\label{Eq:TurningRadius}
 \end{equation}
and decreases with increasing $l$.
$\omega$ in the equation is the wave's temporal frequency.
Therefore, as we go towards the center of the Sun, we only need to resolve waves of lesses and lesses spherical harmonic degrees.
Carrying higher degrees than required would, in fact, unnecessarily limit the time step.
In order to avoid this, we use damping coefficients that are $l$-dependent.
Perturbations of spherical order $l$ do not travel below their corresponding turning radius as given by Eqn.~\ref{Eq:TurningRadius}, and can be damped.
However, we leave about 100 Mm of space below the turning radius before damping starts since the waves have finite extend.
Such damping effectively removes high-$l$ perturbations in the deep interior without effecting the propagation of helioseismic waves.
Modes below $l=40$ are not damped at all because wave of such $l$ can actually travel very close and even through the solar center.

Without flows in the background model, waves can, within the physics of this setup, only gain energy through the acoustic source term, and they can only loose energy through the top boundary.
There is no feedback between waves and the sources.
These simulations run stable and will eventually reach an energy equilibrium.
In the the presence of mass flows in the base state, in particular with strong velocity gradients, it is conceivable that waves may gain momentum and energy from wave-flow instabilities, such as Kelvin-Helmholtz instability.
Such growth would be proportional to the wave's own momentum.
Our simulations have shown however that for solar-type flows such instabilities grow very slowly, and adding a small amount of viscous damping (not shown in the above equations) is enough to ensure long-term stability.

\subsection{Flow Model and Simulation Run}

\begin{figure}[ht]
	\centering
	\hspace*{-0.6cm}\includegraphics[width=\linewidth]{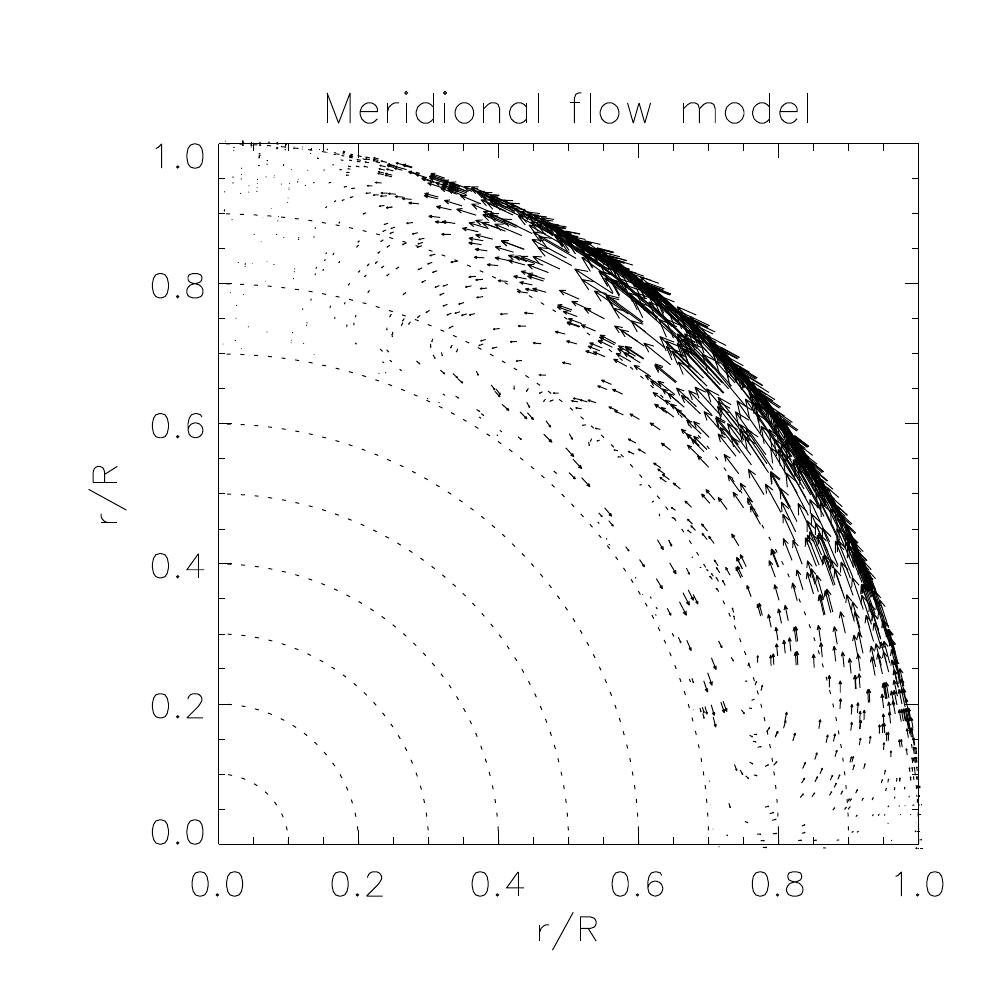}
	\caption{\footnotesize
	Visualization of the meridional flow model of~\citet{2006ApJ...647..662R} used in the simulation. The arrows indicate the direction of the flow, and their size is proportional to the flow speed. The Sun's rotation axis is at the left edge of the panel with the north pole at the top. The flow in the southern hemisphere (not shown) is given by mirror-symmetry about the equatorial plane.
	}
 	 \label{FlowModel}
\end{figure}

For this study, we have simulated the propagation of helioseismic waves through a stationary model of the solar meridional circulation.
The original flow model is the ``reference model'' from~\citet{2006ApJ...647..662R}.
A visualization of the model is shown in figure~\ref{FlowModel}.
This meridional flow is characterized by a single-cell circulation in each hemisphere with a deep return flow that starts at a depth of about 150 Mm below the surface.
The model has a maximum poleward velocity near the surface of approximately 14 m/s and a maximum return flow velocity of 3 m/s.
Detecting such weak flows from helioseismic measurements is extremely challenging due to the inherent randomness of solar oscillations resulting in small signal-to-noise ratios (S/N).
It is expected that very long observations on the order of many years are needed to measure the flow in the deepest parts of the convection zone~\citep{2008ApJ...689L.161B}.
Unfortunately, it is not practical to simulate such long time series giving current computer capabilities.
The present simulation required approximately 1 day of computing time for every three hours of solar time using 264 cores on the {\em Pleiades} supercomputer at the National Aeronautics and Space Administration's (NASA) Advanced Supercomputing Division (NAS). 
However, the most measured quantities in helioseismology -- in particular apparent travel-times, frequency shifts, etc -- depend linearly on the causing perturbations in the solar interior, i.e. sound speed variations or flows.
This is a true as long as the perturbations are small, i.e. in the case of flows: as long as the flow velocities are small compared to the speed of sound.
Therefore, we can improve the S/N without changing the physics by uniformly increasing the amplitude of the background flow model in the simulation.
Such a simulation run for a manageable time is in some sense equivalent to a simulation of the original model but run much longer -- thanks to T.L.~Duvall, Jr. (2011, private communication) for this suggestion. 
In this work, we have increased the meridional flow by a factor of approximately 36 such that the maximum flow velocity is 500 m/s.
This speed is still significantly smaller than the sound speed anywhere in the solar interior, but provides a large increase in S/N.
In fact, since the (uncorrelated) realization noise reduces approximately with the square root of the length of the time series, our simulation should have a S/N that is equivalent to a simulation of the original model but run for a period $36^2$~times longer.

In the following, we list other parameters of the simulation in the study.
The simulation resolves spherical harmonic degrees from 0 to 170.
It is sufficient to capture only this range of small to moderate spherical harmonic degrees because we are only interested in large scale and deep flows.
The time step was 1 second; results were saved with a cadence of 30 seconds; and the simulation produced a total of approximately 76 hours of data.
The first 4 hours of data where discarded because they represent transient behavior as the simulation was started from a model without any waves in it. 
In the end, a total of 4,096 minutes of data have been produced and analyzed.

\section{Measurement technique}
\label{Sec:Measurement}

\begin{figure}[ht]
	\centering
	\hspace*{-1cm}\includegraphics[width=1.15\linewidth]{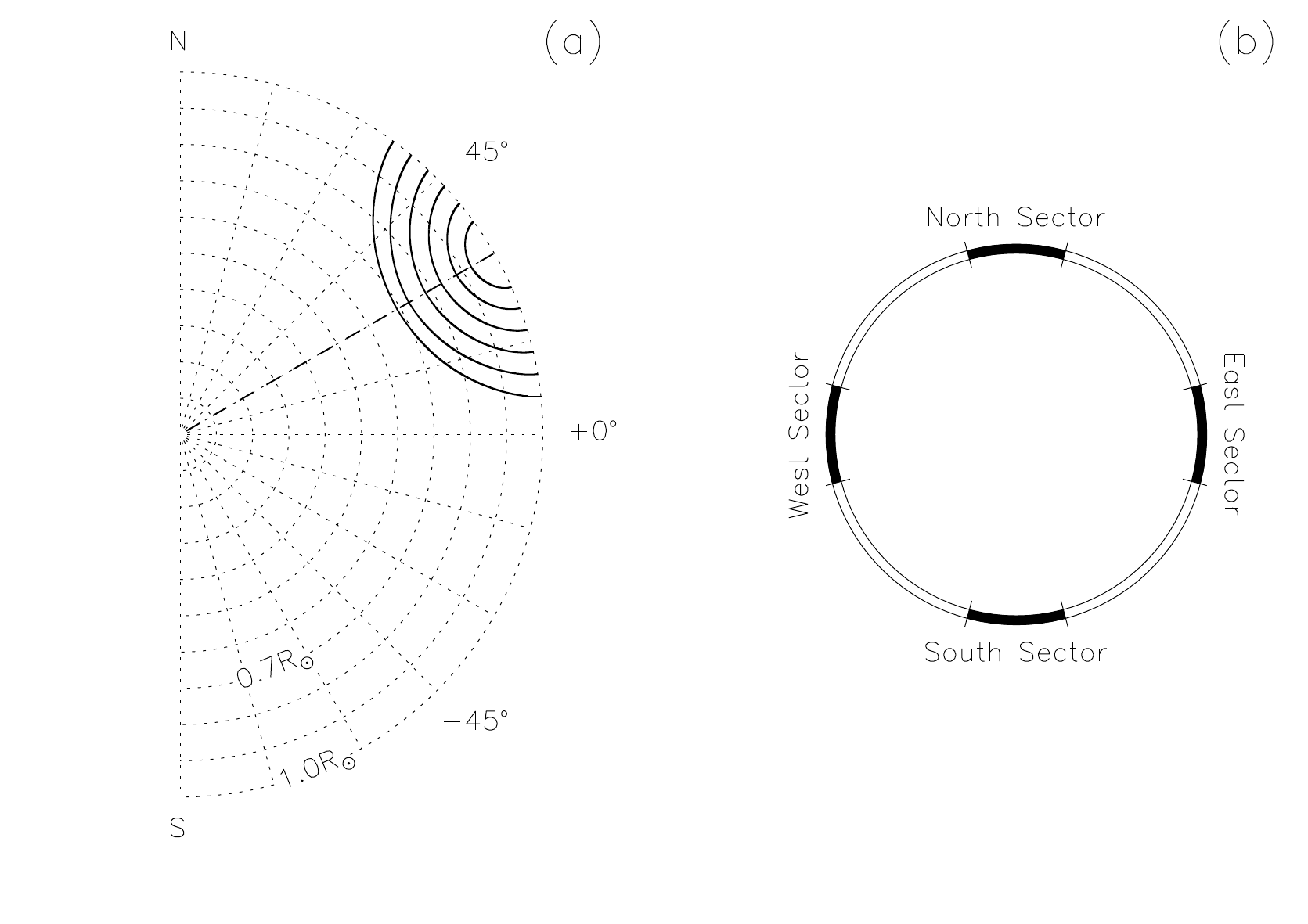}
	\caption{\footnotesize 
	Visualization of the deep focusing measurement scheme used in this study. Panel ({\it a\/}) shows examples of acoustic ray paths ({\it solid lines\/}) originating from and traveling to a range of annuli centered around a latitude of 30\degree. Panel ({\it b\/}) shows a measurement annulus and its decomposition into sectors.
	}
 	 \label{MeasurementScheme}
\end{figure}

In this study, we want to measure the effects that the meridional flow in the simulated Sun has on acoustic waves.
When waves are advected by a flow, their travel time is reduced when they travel in the same direction as the flow and their travel time increases when they travel against the flow.
The travel times are also effected by localized sound speed variations, however, the difference between the travel times of waves going in opposite directions along the same travel path is to first order not sensitive to local sound speed variations because both directions are effected the same way by such perturbations. 
The difference is only sensitive to the flow along the travel path. 

In the following we describe the scheme we used to measure the apparent acoustic travel-time differences between waves traveling northward and waves traveling southward, in the following referred to as N-S travel-time difference, as well as the travel-time difference between east- and west-traveling waves (E-W travel-time difference).
This N-S difference is sensitive to the meridional flow.

From the simulation, we used a 4,096-minute-long time series of the radial component of the perturbation velocity at a fixed geometric height of 300 km above the photosphere.
The simulation provides data on the full solar surface, however, we processed these data more similar to how one would process observations.
First, the time series is split in 1,024-minute-long segments with 50\% overlap.
In each segment, we select $120\degree \times 120\degree$ tiles centered at 0\degree, 90\degree, 180\degree, and 270\degree\, in longitude and -30\degree, 0\degree, and +30\degree\, in latitude, and remap these into heliographic coordinates using Postel's projection with a pixel size of 0.6\degree, the same resolution as {\it SOHO\/}/MDI medium-\cal{l} data~\citep{1997SoPh..170...43K}. 

The geometry of the measurement scheme is the following:
On the surface, centered around each longitude and latitude, we select a series of one-pixel-wide annuli ranging in diameter from 12\degree\, to 42\degree\, in 1.2\degree\, increments.
Examples of ray paths traveling from a point on an annulus to an opposing point is shown in Figure~\ref{MeasurementScheme}a.
In each annulus, the instantaneous signal is averaged over 30\degree-wide sectors in the north, south, east, and west direction (Fig.~\ref{MeasurementScheme}b).
We then cross-correlate the sector-averaged signal with that from its opposing sector, i.e. the north sector with the south sector and the west sector with the east sector.
Cross-correlations for the same latitude within each tile are averaged together.
The longitude-averaged cross-correlation functions for positive and negative time-lag are then fitted separately using Gabor wavelets \citep{1997ASSL..225..241K}, and the resulting phase-travel times are subtracted from each other yielding the apparent travel-time difference between north- and south-going (N-S) waves and between east- and west-going (E-W) waves.
A very small fraction of far outliers with travel-time differences larger than 2 minutes are misfittings, and are discarded.
Measurement schemes of this type have previously been called deep-focusing schemes \citep[e.g.,][]{2009ApJ...702.1150Z} but should not be confused with other deep-focusing schemes such as \citet{1995ASPC...76..465D} which are constructed such that ray paths cross or ``focus'' at a target location below the surface.
\citet{2009ApJ...702.1150Z} used a scheme very similar to the present one to measure deep sound speed perturbations.
However, they employed 90\degree-wide measurements sectors, i.e. quadrants, instead of the 30\degree-wide sectors used here.
 
\section{Results}
\label{Sec:Results}

\begin{figure*}[ht]
	\centering
	\hspace*{-0.2cm}\includegraphics[width=\linewidth]{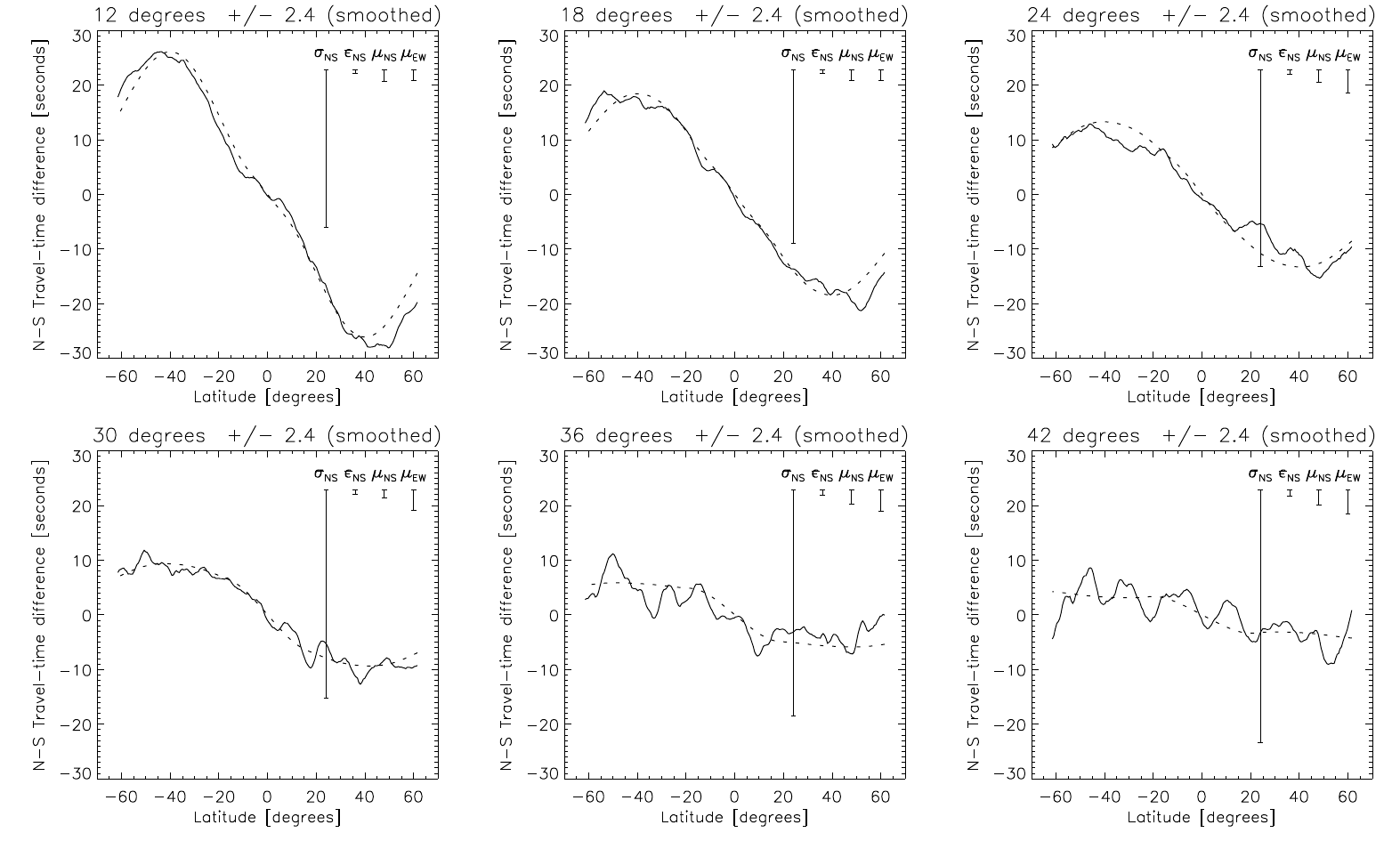}
	\caption{\footnotesize 
	Travel-time differences between north-going and south-going waves for 6 different travel distances of 12, 18, 24, 30, 36, and 42 heliographic degrees. The panels show the values measured from 4,096 minutes of simulation data ({\it solid curves\/}) and corresponding ray-theory calculations ({\it dotted curves\/}). All values have been averaged over a range of $\pm 2.4\degree$ in travel distance and smoothed over $\pm 3\degree$ in latitude. Error bars in each panel show the size of the latitudinal average of the standard deviation, $\overline\sigma_{NS}$, a measure of the scatter between individual measurements, and the standard error of the mean, $\overline\epsilon_{NS}$, as defined by Equations~\ref{SigmaEquation} and~\ref{EpsilonEquation}, respectively. Also, measures of the deviation of the measured travel-time diffrences from the their ray-approximation values are show by $\mu_{NS}$ for the N-S travel-time differences plotted here, and for comparison the corresponding $\mu_{EW}$ for the E-W travel-time differences from Figure~\ref{EWTravelTimeDifferences}.
	}
 	 \label{NSTravelTimeDifferences}
\end{figure*}
\begin{figure*}[ht]
	\centering
	\hspace*{-0.2cm}\includegraphics[width=\linewidth]{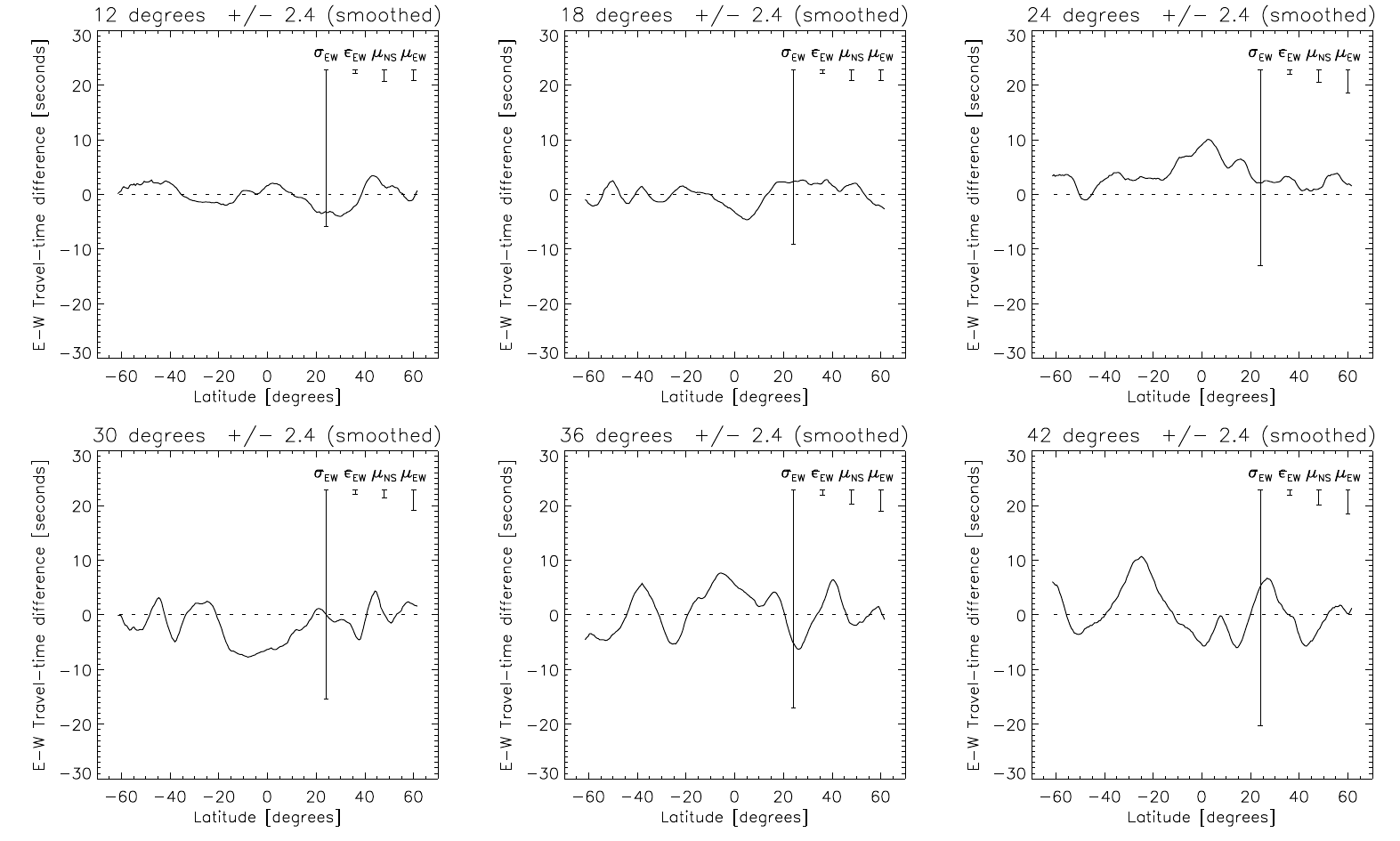}
	\caption{\footnotesize 
	Travel-time differences between east-going and west-going waves ({\it solid curves\/}) computed from the same data and for the same travel distances as in Figure~\ref{NSTravelTimeDifferences}.  Values have also been averaged over a range of $\pm 2.4\degree$ in travel distance and smoothed over $\pm$ 3\degree\, in latitude. Since no flow in east-west direction was prescribed in the simulation, the travel time differences should be zero ({\it dotted curves\/}) if there was no measurement noise. Given error bars are analogous to those in Figure~\ref{NSTravelTimeDifferences}.	}
 	 \label{EWTravelTimeDifferences}
\end{figure*}

\subsection{Travel-time differences}
\label{Sec:TTD}

N-S travel-time differences measured from the simulation data for a range of distances are presented in Figure~\ref{NSTravelTimeDifferences}.
The travel distances of $12\degree$, $18\degree$, $24\degree$, $30\degree$, $36\degree$, and $42\degree$ correspond to lower turning points of the ray paths of approximately 52, 78, 106, 134, 164, and 195~Mm below the photosphere, respectively.
Since individual measurements are fairly noisy, it is advantageous to trade some spatial resolution for reduced noise.
The results have therefore been averaged over a small range of travel distances ($\pm 2.4\degree$) as well as a range of latitudes ($\pm 3\degree$).

Figure~\ref{NSTravelTimeDifferences} also shows the N-S travel-time differences expected from ray-approximation calculations for the prescribed flow.
Assuming small perturbation, the ray paths are the same as in the case without flows, and the travel-time difference can be computed by integrating along the ray path the flow component tangential to the ray. 
Specifically, in the first approximation the travel-time difference between waves going along a ray path and going along the same path in opposite direction is given by:
\begin{equation}
   \Delta t = - 2 \int_{ray path} \frac{\mathbf{v_0}\cdot\mathbf{T}}{c_0^2} ds
\end{equation}
where $\mathbf{T}$ is the unit vector tangential to the ray path \citep[e.g.,][]{1997ASSL..225..241K}.
The calculations in this study used ray paths computed from the ray-tracing code of \citet{2009SoPh..257..217C}. 

We can see from Figure~\ref{NSTravelTimeDifferences} that the results from the analysis of the simulated data are very close to the expected travel-time differences computed using ray theory.
This is true especially for the four smaller travel distances.
The signal-to-noise ratio (S/N) is high here since the deviations from the ray-approximation calculations are rather small.
For the two larger distances, however, the noise seems to be comparable with the signal.

\subsection{Error estimation}
\label{Sec:Error}

When working with measurements, whether using actual observations or artificial data from numerical simulations, it is imperative to estimate the accuracy of the measurement results. 
This has been an important issue in helioseismology where the results are obtained using complicated data analysis procedures. 
Therefore, we present the error estimation in detail. 
Several error estimates are shown in Figure~\ref{NSTravelTimeDifferences} and are explained in the following.
Each data point in Figure~\ref{NSTravelTimeDifferences} is the mean of a large number of individual measurements -- these are the travel-time differences from individual fittings of the cross-correlation function for the different measurement tiles, a range of distances, and a range of latitudes.
From these, we can compute a statistical error estimate.
Let us denote the individual measurements by $y_i$ with $i \in [1,N]$, where N is the number of individual measurements averaged for each final data point, $\bar{y}$.
Here:
\begin{equation}
  \bar{y}= \frac{1}{N} \displaystyle\sum_{i=1}^{N} y_i.
\end{equation}
The scatter in the individual measurements is given by the standard deviation, $\sigma_{NS}$, defined as:
\begin{equation}
  \sigma_{NS} = \sqrt{ \frac{1}{N} \displaystyle\sum_{i=1}^{N} (y_i-\bar{y})^2 }.
  \label{SigmaEquation}
\end{equation}
$\sigma_{NS}$ is a measure of the deviation of an individual measurement from the mean $\bar{y}$.
However, we are not really interested in the error of an individual measurement but rather in the accuracy of the mean $\bar{y}$.
Its error can be estimated by
\begin{equation}
   \epsilon_{NS} = \frac{\sigma_{NS}}{\sqrt{N}}
   \label{EpsilonEquation}
\end{equation}
if we can assume statistical independence of the individual measurements.
This estimate, $\epsilon_{NS}$, is often called the standard error of the mean.
As can be seen from Figure~\ref{NSTravelTimeDifferences}, $\epsilon_{NS}$ is very small and is in fact significantly smaller than the difference between the measurements and the ray-approximation calculations.
This indicates that there are either systematic differences between the measurements and the ray-approximation calculations or that statistical independence is a poor assumption.

An independent error estimation can be derived by looking at the deviation of the E-W travel-time differences from their expected values.
It seems reasonable to assume that the error in both directions is of the same order.
Results for the E-W travel-time difference are shown in Figure~\ref{EWTravelTimeDifferences}.
Since no azimuthal flow is prescribed in the simulation, E-W travel-time differences should strictly vanish for all latitudes without using any approximation, such as the ray approximation.
However, due to the finite length of the measurement, non-zero values are found.
The deviations can be used as an estimate of the error in both E-W and N-S travel-time differences.
We define:
\begin{equation}
   \mu_{EW}=  \sqrt{ \frac{1}{M} \displaystyle\sum_{j=1}^{M} \left[\bar{y}(\theta_j) - \bar{y}_0(\theta_j)\right]^2 }
   \label{MuEquation}
\end{equation}
where $\bar{y}(\theta_j)$ is now the mean E-W travel-time difference as a function colatitude $\theta_j$, M is the number of different colatitudes, and $\bar{y}_0(\theta_j)$ is the theoretically expected travel-time difference and is zero in this case.
As is evident from Figures~\ref{NSTravelTimeDifferences} and \ref{EWTravelTimeDifferences}, $\mu_{EW}$ seems to be a more appropriate estimate of the noise in both the N-S and E-W travel-time measurements than $\sigma_{NS}$ and $\sigma_{EW}$.
Also shown in these figures are values of the deviation of the N-S travel-time difference from their ray-approximation values, i.e. $\mu_{NS}$.
It is computed like equation~\ref{MuEquation} except that $\bar{y}_0$ is replaced by the N-S travel-time difference computed from the ray approximation.
Values of $\mu_{NS}$ and $\mu_{EW}$ are similar indicating that the systematic error made by the ray approximation is probably small compared to the statistical variability of the measurements.
It is somewhat surprising that $\mu_{EW}$ is actually slightly larger than $\mu_{NS}$ for the longest travel distances considered here.
However, the scatter in the N-S and E-W travel-time measurements, i.e. $\sigma_{NS}$ and $\sigma_{EW}$, are very much the same as seen in Figures~\ref{NSTravelTimeDifferences} and~\ref{EWTravelTimeDifferences}.
 
\subsection{Signal-to-noise ratio}
\label{Sec:SNR}

\begin{figure}[]
	\centering
	\hspace*{-0.4cm}\includegraphics[width=0.95\linewidth]{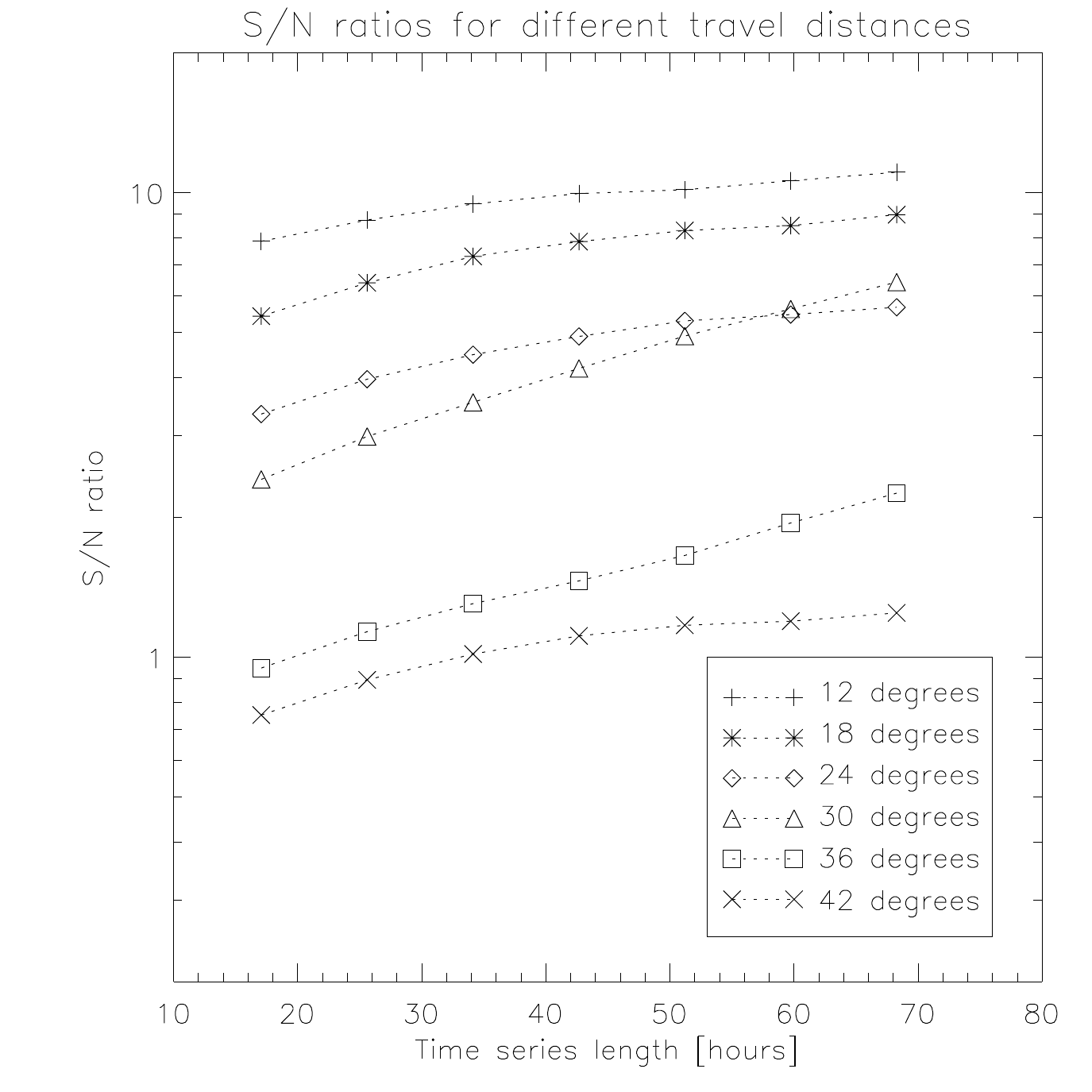}
	\caption{\footnotesize 
	S/N of the N-S travel-time difference measurements as a function of the length of the analyzed time series for different travel distances. S/N is here defined as the ratio between the maximum amplitude over latitudes of the ray-approximation N-S travel-time difference and the standard deviation, $\mu_{NS}$, of the measured travel-time differences from their ray-approximation values.
	}
 	 \label{SNRatios}
\end{figure}
\begin{figure}[]
	\centering
	\hspace*{-0.2cm}\includegraphics[width=0.90\linewidth]{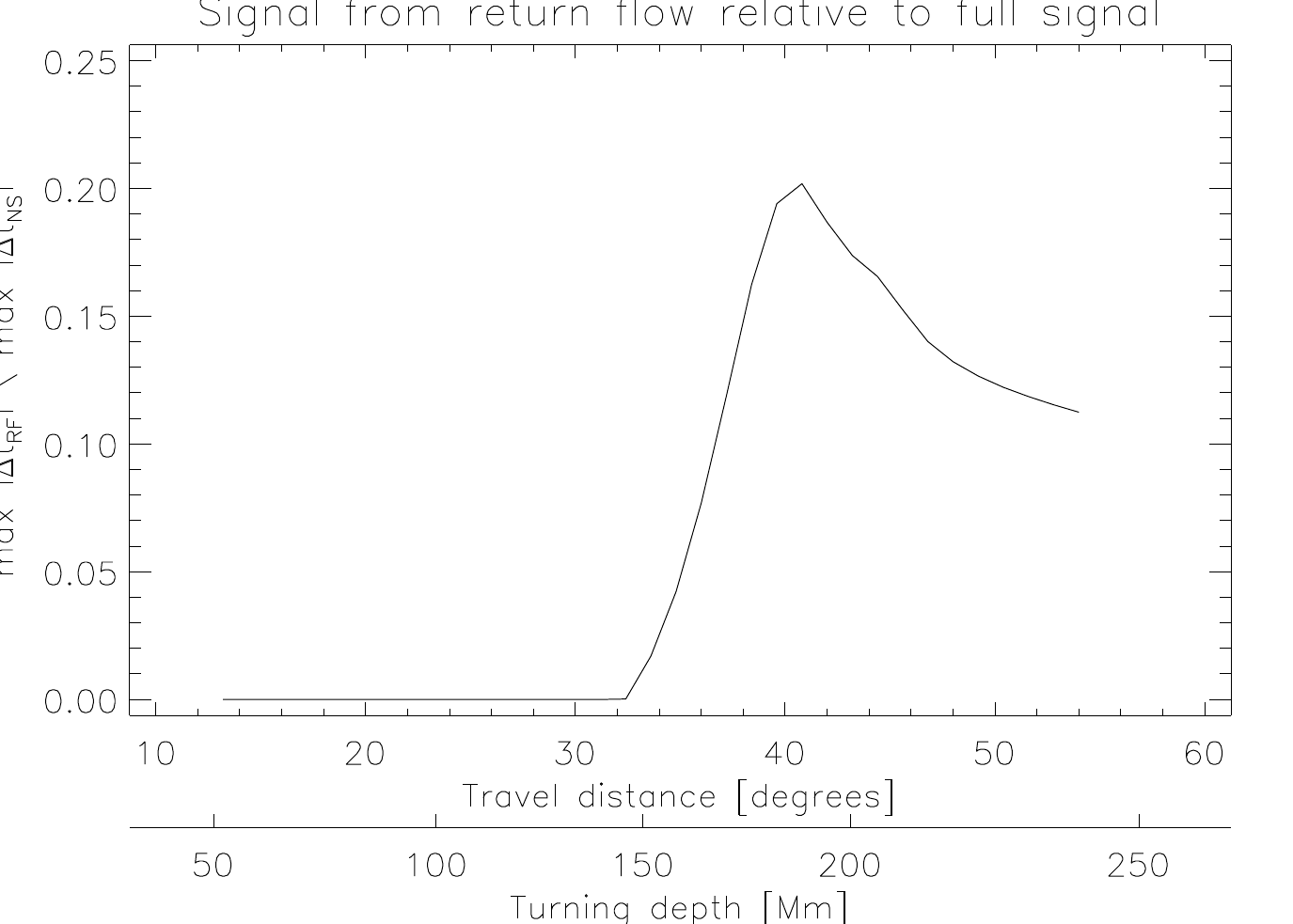}
	\caption{\footnotesize 
	Ratio between the amplitudes of the N-S travel-time differences from ray approximation computed for the return flow alone, $\Delta t_{RF}$, and for the full meridional flow, $\Delta t_{NS}$, as a function of travel distance. The peak contribution from the return flow in this case corresponds to about 0.9 seconds in travel-time difference.
	}
 	 \label{RatioReturnFlow}
\end{figure}

Using the amplitude of the ray-approximation N-S travel-time difference and an estimates of the measurement noise in the form of the deviations $\mu_{NS}$, we can compute a signal-to-noise ratio (S/N) for our measurements.
The results are shown in Figure~\ref{SNRatios} for a range of measurement lengths.
As one would expect, S/N increases approximately as the square root of the measurement time.
We can use this dependency to roughly estimate the duration of a time series one would need to measure, say, the flow near the bottom of the convection zone to a certain accuracy.
For a heliographic distance of 42\degree\, and 72 hours of measurement time (3 days) we have a S/N of approximately 1.25 according to Figure~\ref{SNRatios}.
However, remember that the flow velocities in the present simulation have been increased from their realistic values to 500 m/s at the surface to make the simulation and measurements feasible.
For a more realistic meridional flow of, say, 20 m/s, the signal-to-noise ratio would be 25 times smaller.
Assuming a S/N of 2 is desired, one would then need $(25 \times 2 / 1.25)^2 \times 3$ days or on the order of 10 years.
MDI medium-l measurements are available for almost continuous 15 year, so this may be already possible.
However, note that the signal for such long measurement distances still remains dominated by the strong poleward flow in the upper layers of the Sun.
As shown in Figure~\ref{RatioReturnFlow}, only a small portion of the signal -- up to about 20\% at its peak -- is from the return flow which in the model considered here starts at a depth of approximately 146 Mm.
So, a higher signal-to-noise ration may be desired.

\section{Conclusions}

We have simulated the propagation of acoustic waves in the full solar interior in the presence of a prescribed meridional flow with a deep return flow, and we performed time-distance helioseismology measurements to detect the effects of the meridional circulation on the acoustic travel-times difference between north- and south-going waves. 
The measurements were done for large travel distances between 12 and 42 heliographic degrees corresponding to lower turning points of the acoustic waves between 52 and 195~Mm below the photosphere, i.e. deep in the solar convection zone all the way to the tachocline.
The flow velocity in the model was artificially increased by a significant factor to a value of 500 m/s in order to model the flow measurements using relatively short time series that can be calculated on currently available supercomputer systems.
The results show that this approach works well without significantly changing the physics of wave propagation, as expected from theoretical grounds.
The results also show that it is, in fact, possible to measure the effects of a meridional flow in the deeper solar convection zone by employing a deep-focusing time-distance helioseismology technique.
Within the statistical variability (noise) of the measurements, the measured N-S travel-time differences agree well with the ray-approximation calculations.
For distances between $12\degree$ and $30\degree$ corresponding to lower turning depths between 52 and 136~Mm, the agreement is in fact excellent, and still good for $36\degree$ (164 Mm depth).
Noise starts to dominate for the very longest travel distance, however.
We estimate that for realistic values of the meridional flow velocity $\sim10$ year time-series or longer may be needed for adequate S/N.
Such data are currently available from the SOHO and SDO space observatories (since 1996), and ground-based GONG network (since 1996).
It should be mentioned that the present simulation uses rather simple models for the excitation of acoustic waves as well as wave damping, and that therefore the noise properties of the Sun may not be very accurately represented in this numerical model.
None-the-less does it seem clear than very long helioseismology observations are needed in order to detect small flows at the base of the convection zone.
Still, S/N may be increased, for instance, by the use of phase-speed filters which we have not explored here, or by more spatial averaging.
We also leave it for future work to develop and perform an inversion to infer actual flow velocities from the measured travel-time difference.
It seems, however, that the current ray-approximation based travel-time inversion techniques are sufficiently accurate.

%

%

%
\bibliographystyle{apj}
\bibliography{hartlep_bibliography_archive}  

\end{document}